\documentclass[aps,prl,epsfigure,twocolumn]{revtex4}
\usepackage{amsmath}    
\usepackage{epsfig}
\usepackage{epstopdf}
\usepackage{times}

\newcommand{\miniket}[1]{\vert#1\rangle}
\newcommand{\id}{\;\mbox{$\rm{I} \hspace{-2.0mm} {\bf I}$}\,}
\newcommand{\minibra}[1]{\langle#1\vert}
\newcommand{\minisand}[3]{\langle#1\vert#2\vert#3\rangle}

\begin{document}

\title{Optimal path for a quantum teleportation protocol in entangled networks}

\author{C. Di Franco$^{1,2}$ and D. Ballester$^2$}

\affiliation{$^1$ Department of Physics, University College Cork, Cork, Republic of Ireland\\
$^2$ Departamento de Qu\'imica F\'isica, Universidad del Pa\'is Vasco-Euskal Herriko Unibertsitatea, Apartado 644, E-48080 Bilbao, Spain}

\begin{abstract}
Bellman's optimality principle has been of enormous importance in the development of whole branches of applied mathematics, computer science, optimal control theory, economics, decision making, and classical physics. Examples are numerous: dynamic programming, Markov chains, stochastic dynamics, calculus of variations, and the brachistochrone problem. Here we show that Bellman's optimality principle is violated in a teleportation problem on a quantum network. This implies that finding the optimal fidelity route for teleporting a quantum state between two distant nodes on a quantum network with bi-partite entanglement will be a tough problem and will require further investigation.
\end{abstract}

\maketitle

Finding the route between two nodes of a given graph such that the sum of the weights associated with the links within this path is minimized is arguably one of the most fundamental problems in graph theory~\cite{diestel}. Decades ago, an algorithm to solve it was proposed by Dijkstra~\cite{dijkstra}. Its applicability relies on the ability to compute the optimal path for a large-scale network based on the optimization performed at smaller parts of it, with initial conditions given by other parts. As stated by Dijkstra, to find the path of minimal total length between two given nodes $P$ and $Q$, ``we use the fact that, if $R$ is a node on the minimal path from $P$ to $Q$, knowledge of the latter implies the knowledge of the optimal path from $P$ to $R$''. This can be understood as a particular instance of the more general optimality principle by Bellman~\cite{bellman}: for problems satisfying this principle, {\it the global optimal solution can be determined in terms of local optimal ones for smaller subproblems (optimal substructure)}. This property has been extremely useful in the study of dynamic programming, control theory, economics, and Markovian stochastic processes.

Several problems originally thought in classical scenarios have been subsequently redefined and studied when dealing with quantum systems. Significant differences in the predictions are found when focusing on atomic and sub-atomic scales, but even some macroscopic phenomena can be understood only through a complete analysis by means of quantum theory. In particular, the role of entanglement~\cite{horodecki} in different processes in nature is currently under deep investigation~\cite{natureentanglement}. On the other hand, the generalization of classical information theory to quantum scenarios has paved the way to the development of quantum information theory, with rules that are fundamentally different from classical ones~\cite{nielsen}.

Here we focus our attention on a particular scheme of quantum information theory, known as quantum teleportation, that has proved to be one of the most striking applications of entanglement as a resource~\cite{teleportation}. By means of a shared entangled channel and a conditional local operation, a receiver (Bob) is able to reconstruct the unknown state of a qubit given to a sender (Alice), after she performs a joint measurement and communicates classically her result to him. Considerable effort has been made in demonstrating quantum teleportation experimentally by means of polarized entangled photons~\cite{zeilingerandboschiteleportation}, squeezed-state entanglement~\cite{polzikteleportation}, liquid-state nuclear magnetic resonance~\cite{laflammeteleportation}, and trapped ions~\cite{blattteleportation}. {\it In this paper, we show that Bellman's optimality principle is violated when we consider the teleportation protocol on a quantum network}.

Although further generalizations and extensions of Dijkstra's algorithm have been studied~\cite{cormen}, here we will shortly illustrate its working principles. If we want to find the shortest path (in terms of the sum of the weights) between two nodes $A$ and $B$ of a graph, we need first of all to assign a number to each node. We give the value $0$ to $A$ and $+\infty$ to all the others. Then we start from $A$ (which we define as the ``incoming'' node, until it is marked) and we consider its neighboring nodes. To each of them, we assign the minimum between its current value (in this case $+\infty$) and the sum of the value of the incoming node (in this case $0$) and the weight of the link between the latter and the node under investigation. After we have done this for all the neighboring nodes, we mark $A$, and we no longer need to consider it. We then focus our attention on the unmarked node in the network with the smallest number, and we highlight the link that ``contributed'' to generate its actual value. We proceed in the same way with this new unmarked node taking the role of $A$ in the previous description. The process stops when there is a highlighted link ending on $B$. The only highlighted path going from $A$ to $B$ is the shortest path between these two nodes of the graph.

\begin{figure}[t]
\psfig{figure=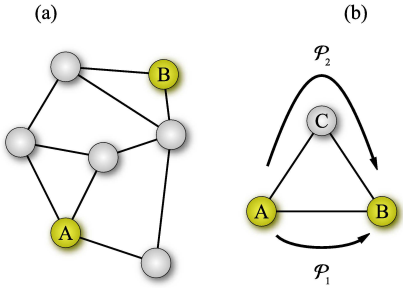,width=5.5cm}
\caption{{\bf (a)} Sketch of the general setting considered: Alice and Bob are connected through a network in which each node shares entangled channels with others. {\bf (b)} Simplified scheme: Alice can teleport her state to Bob's position either by means of the direct link $A-B$ (path ${\cal P}_1$) or through the link $A-C$ and then $C-B$ (path ${\cal P}_2$).}
\label{quantumnetworksketch}
\end{figure}

The investigation on quantum teleportation has been mainly focused on the case where Alice and Bob directly share an entangled resource. Interesting results have been obtained when a sequence of teleportations has to be performed along a chain of nodes~\cite{modlawska}. In order to provide a broader scenario for quantum teleportation, we consider here a more complex setting: Alice and Bob are connected through a network in which each node shares entangled channels (that are, in general, non-maximally entangled) with others. A sketch of the situation under investigation is presented in Fig.~\ref{quantumnetworksketch}{\bf (a)}.
This is in line with the recent interest that arose in the scientific community about the possibility to realize quantum networks~\cite{networks}. Networking distant nodes is a fundamental step in designing and building distributed quantum computers, as well as in the implementation of large-scale highly secure quantum communication protocols. This concept paved the way for the promising idea of quantum internet~\cite{internet}. We focus, in particular, on quantum networks with just bi-partite entanglement (therefore, there is no multi-partite entanglement in our network).

The scenario described here resembles the shortest-path problem aforementioned. Surprisingly, we will find that Dijkstra's algorithm can be adapted for solving our problem, while the entangled channels are pure. When the nodes in our network are linked by means of mixed resources, we can still consider the analogy with the graph theory problem, but in general we can no longer use Dijkstra's method. {\it Moreover, we will find that our problem does not satisfy the broader Bellman's principle and then also all the other algorithms based on it cannot be used}. We will discuss how to determine for which cases the principle is still satisfied.

Let us start considering a simple scenario, where Alice, Bob, and Charlie (an agent located in a third node of the network) are linked by means of non-maximally entangled pure states, as shown in Fig.~\ref{quantumnetworksketch}{\bf (b)}. The extension to more complex network structures and mixed channels is discussed later. We want to stress again that we are dealing with the case of bi-partite entangled resources here. This assumption, together with the others within this paper, is a case that can be experimentally tested with current state-of-the-art quantum technologies. A general non-maximally entangled pure state of two qubits can be cast, in an appropriate basis, as $\miniket{\Phi^{(\theta)}}_{1,2}=\cos\theta\,\miniket{0}_1\miniket{0}_2+\sin\theta\,\miniket{1}_1\miniket{1}_2$, where $\theta \in [0,\pi/4]$ is a parameter depending on the amount of entanglement present in the channel (related to Schmidt coefficients)~\cite{preskill}. As a measure of entanglement, we will use the negativity~\cite{negativity} of the density matrix $\rho_{1,2}$ describing the state of qubits $1$ and $2$ (in the case of pure states, this corresponds to $\rho_{1,2}=\miniket{\Phi^{(\theta)}}_{1,2}\minibra{\Phi^{(\theta)}}$). In the scenario considered so far, we have that the negativity ${\cal N}_{1,2}$ of the state $\miniket{\Phi^{(\theta)}_{1,2}}$ is simply ${\cal N}_{1,2}=\sin{2\theta}$.

Alice has two possible choices to teleport her input state to Bob's position (we assume that the standard teleportation protocol is used at each step): either she can use the direct link $A-B$ [path ${\cal P}_1$ in Fig.~\ref{quantumnetworksketch}{\bf (b)}], or she can teleport her input state to Charlie through the link $A-C$, and he will subsequently teleport it to Bob through the link $C-B$ [path ${\cal P}_2$ in Fig.~\ref{quantumnetworksketch}{\bf (b)}]. Let us define the negativity of the three shared entangled states, corresponding to these three links, as ${\cal N}_{A,B}$, ${\cal N}_{A,C}$ and ${\cal N}_{C,B}$, respectively. The result of a deterministic teleportation protocol (we are not dealing with probabilistic schemes~\cite{probabilistic} in this paper) is the final state $\rho^{(fin)}_B$ at Bob's location. This is, in general, a mixed state that is not exactly the same as the initial one $\rho^{(in)}_A=\miniket{\psi}_A\minibra{\psi}$ that Alice wanted to teleport. Although we consider here a pure input state, the extension to the mixed case is straightforward. Moreover, we focus the investigation first on azimuthal states of the form $\alpha\miniket{0}+\beta\miniket{1}$, with $|\alpha|^2+|\beta|^2=1$ and $\alpha$, $\beta$ both real~\cite{crapdaniel,crapbob}.

In order to evaluate the performance of the scheme, we need to estimate how close $\rho^{(fin)}_B$ is to $\rho^{(in)}_A$ by means of the teleportation fidelity ${\cal F}={}_A\minisand{\psi}{\rho^{(fin)}_{B}}{\psi}_A$. This value clearly depends on the state to transfer; therefore an average over all the possible input states is in order. Let us define $\bar{\cal F}$ as the average of ${\cal F}$ over all the possible $\miniket{\psi}_A$ uniformly distributed. Interestingly, the average fidelity if Alice performs a direct teleportation through the link $A-B$ can be simply cast as $\bar{\cal F}_{A-B}=(3+{\cal N}_{A,B})/4$. On the other hand, if she chooses to use the two-link path $A-C-B$, the fidelity is $\bar{\cal F}_{A-C-B}=(3+{\cal N}_{A,C}{\cal N}_{C,B})/4$ (a similar result has been obtained in Ref.~\cite{rigolin}). Clearly, the general case through a more complex path ${\cal P}$ can be written as $\bar{\cal F}_{\cal P}=(3+\prod_{i \in {\cal P}}{\cal N}_{i})/4$, where ${\cal N}_i$ denotes the negativity of the entangled channel corresponding to the link $i$ within the path ${\cal P}$. Here it is easy to note how to map the maximization of the teleportation fidelity to the aforementioned shortest-path problem. If we consider the value $-\ln {\cal N}_i$ ($0\le{\cal N}_i\le1$, so we have $-\ln {\cal N}_i\ge0$)  as the weight of link $i$, the maximization of $\bar{\cal F}_{\cal P}$ corresponds to the shortest path solution ({\it i.e.}, to finding the path ${\cal P}$ that minimizes $-\sum_{i \in {\cal P}}\ln {\cal N}_{i}$).

Let us now extend the analysis to the case where the shared resources are not pure. We first start by describing the main result, obtained for the interesting class of $X$-states~\cite{xstates}. This includes, among others, maximally entangled Bell states and Werner states~\cite{werner}.  This class is relevant in many physical settings, and, for this reason, it has recently attracted the attention of the scientific community and has been investigated theoretically as well as experimentally~\cite{xstatespapers}. An $X$-state of two qubits is described by the density matrix
\begin{equation}
\rho=
\begin{pmatrix}
a_{11}&0&0&a_{14}\\
0&a_{22}&a_{23}&0\\
0&a_{32}&a_{33}&0\\
a_{41}&0&0&a_{44}
\end{pmatrix},
\end{equation}
with $a_{jk}=a_{kj}^*$ ($a_{kj}^*$ denotes the complex conjugate of $a_{kj}$). Following our discussion, we study the average fidelity when Alice teleports her input state through a general path. Straightforward but cumbersome calculations lead to
\begin{equation}
\label{fidelitymixed}
\begin{split}
\bar{\cal F}_{\cal P}=\frac{1}{4}[&2+\prod_{i \in {\cal P}}(a_{11}^{(i)}-a_{22}^{(i)}-a_{33}^{(i)}+a_{44}^{(i)})\\
&+\prod_{i \in {\cal P}}(a_{14}^{(i)}+a_{23}^{(i)}+a_{32}^{(i)}+a_{41}^{(i)})].
\end{split}
\end{equation}
Here $a_{jk}^{(i)}$ is the corresponding element of the density matrix describing the entangled state of link $i$ within the path ${\cal P}$. It is now straightforward to see why, in general, Dijkstra's algorithm can no longer be used to solve the problem of finding the corresponding shortest path: the presence of two different non-constant terms in the formula for $\bar{\cal F}_{\cal P}$ in Eq.~(\ref{fidelitymixed}), each one equal to the product of values associated with the links in the path, does not allow us to assign to the links a single weight with the additive property. {\it We can show more neatly the violation of Bellman's optimality principle} with the following example: assume that, by applying the fidelity formula in Eq.~(\ref{fidelitymixed}) for the case of $X$-states, we determine that ${\cal P}_1$ is the optimal path to go from $A$ to $B$ in Fig.~\ref{quantumnetworksketch}{\bf (b)}. Next, imagine that we add a node $D$ connected only to $B$ through an entangled link of the same class. If we redefine our task as teleporting the unknown state from $A$ to $D$, then it is impossible to guarantee that the optimal path for this new problem will contain the subpath ${\cal P}_1$ (but, according to Dijkstra's protocol, if the optimal path from $A$ to $D$ contains ${\cal P}_2$ and not ${\cal P}_1$, then ${\cal P}_1$ cannot be optimal from $A$ to $B$). In other words, {\it the optimization depends not only on the present node and the total ``distance'' accumulated before but also on the whole structure of the network}. This justifies why finding the optimization in this type of networks is not possible by means of any algorithm based on Bellman's optimality principle.

Nevertheless, the problem of maximizing the teleportation fidelity can still be mapped into finding the shortest path in a graph, but this time each link $i$ has two different weights $\mu_i$ and $\nu_i$ and the distance along a path is obtained as $\prod_{i \in {\cal P}}\mu_i+\prod_{i \in {\cal P}}\nu_i+\eta$, where $\eta$ is a constant. The corresponding weights are $\mu_i=a_{11}^{(i)}-a_{22}^{(i)}-a_{33}^{(i)}+a_{44}^{(i)}$ and $\nu_i=a_{14}^{(i)}+a_{23}^{(i)}+a_{32}^{(i)}+a_{41}^{(i)}$, respectively. When $a_{22}^{(i)}=a_{33}^{(i)}=0$, we have $a_{11}^{(i)}+a_{44}^{(i)}=1$ (due to the fact that the trace of a density matrix is always equal to $1$), and Dijkstra's algorithm can be used again. This condition means that the shared entangled state has to be confined in the subspace spanned by $\{\miniket{00},\miniket{11}\}$.

We want to stress here that clearly this result is not a pure quantum effect but, as stated above, it depends on the fact that the teleportation fidelity in the investigated scenario cannot be related to individual link weights with a pure additive or multiplicative property. This can be, of course, also the case in particular problems on classical networks, in which the figure of merit that we want to maximize (or minimize) behaves neither additively nor multiplicatively under the composition of two channels within the network: Bellman's optimality principle is violated, and all the algorithms based on it cannot be used. However, {\it it is interesting that the scenario presented here gives an explicit and intuitive example of this fact in a quantum case}.

Let us now consider a significant example of an $X$-state, a model that can be easily seen as a generalization of two-qubit Werner states~\cite{werner}, in which $\miniket{\Phi^{(\theta)}}$ takes the place of the Bell pair~\cite{crapbob}. In this case, the non-maximally entangled mixed state of qubits $1$ and $2$ is represented by the density matrix $\rho_{1,2}=p_W \miniket{\Phi^{(\theta)}}\minibra{\Phi^{(\theta)}}+(1-p_W)\id^{(4)}/4$, with $p_W \in [0,1]$ and $\id^{(4)}$ being the $4 \times 4$ identity matrix. This can be, for instance, the result of an initially pure general entangled state (represented by $\miniket{\Phi^{(\theta)}}$) that is successively affected by isotropic random noise due to the interaction with the environment~\cite{nielsen}. Let us define the state of qubits $j$ and $k$ as $\rho_{j,k}=p^{(j,k)}_W \miniket{\Phi^{(\theta_{j,k})}}\minibra{\Phi^{(\theta_{j,k})}}+(1-p^{(j,k)}_W)\id^{(4)}/4$, with $\theta_{j,k}$ being the angle in the definition of $\miniket{\Phi^{(\theta_{j,k})}}$. The average teleportation fidelity in this setting is
\begin{equation}
\bar{\cal F}_{\cal P}=\frac{1}{4}[2+\prod_{i \in {\cal P}}p^{(i)}_W+\prod_{i \in {\cal P}}p^{(i)}_W\sin(2\theta_i)].
\end{equation}
This shows that, for this model of mixed resources, Dijkstra's algorithm (and, more generally, Bellman's optimality principle) can no longer be used to solve the problem of finding the shortest path corresponding to the maximization of the average teleportation fidelity.

Now, let us comment on the restriction we have considered on the initial state to be teleported. Clearly, its more general form $\alpha\miniket{0}+\beta\miniket{1}$ is without any condition on $\alpha$ and $\beta$, apart from $|\alpha|^2+|\beta|^2=1$ (they are complex numbers). The set of all the possible initial states over which we have to take the average is thus different. However, if we average over this set, the formulas for the fidelity are cumbersome and cannot be cast in a simple form as for the case investigated above. For the sake of completeness, a longer technical work, including the formulas for a general initial state and further details, is in preparation. However, we want to anticipate here that, also in that case, Bellman's optimality principle is violated for the same conditions as those discussed in this paper.

The formulas presented so far allow one to obtain the maximum teleportation fidelity when the strategy to follow consists only in deciding an optimal route and performing single teleportation steps for each link of the path. However, one can outperform this value of fidelity including a probabilistic preparation stage in the scheme. In contrast to probabilistic teleportation~\cite{probabilistic}, the teleportation proposed here is still deterministic. The probabilistic nature is present only in the preparation stage, in which the network is ``arranged'' in a way so as to achieve the best fidelity. After the preparation stage, one can perform the sequence of teleportations and obtain deterministically Bob's final state. {\it The fidelity will depend on the result of the preparation stage and its minimum value will correspond to the maximum fidelity calculated on the original graph according to the formulas presented in this paper}.

Let us explain the method by considering again the easy instance of network sketched in Fig.~\ref{quantumnetworksketch}{\bf (b)}. Suppose that the shared resources are pure and ${\cal N}_{A-B}>{\cal N}_{A-C}{\cal N}_{C-B}$, with the generalization to more complex network structures and mixed channels being straightforward. According to the previous discussion, the optimal path should be the direct link $A-B$. The entangled resources corresponding to the links $A-C$ and $C-B$ are thus not needed for the implementation of the teleportation protocol in order to achieve the fidelity $\bar{\cal F}_{A-B}=(3+{\cal N}_{A-B})/4$. We can therefore use the couple of entangled pairs $A-C$ and $C-B$ to perform a probabilistic entanglement swapping and concentration scheme in order to create an additional link $A-B$. For instance, by measuring the two qubits at location $C$ on a Bell basis, one can obtain a new entangled resource $A-B$ with negativity ${\cal N}'_{A-B}=2{\cal N}_{A-C}{\cal N}_{C-B}/\gamma$ and a success probability $p=(1-\sqrt{1-{\cal N}_{A-C}}\sqrt{1-{\cal N}_{C-B}})/2$. Here $\gamma=[2-\cos(\delta_1-\delta_2)-\cos(\delta_1+\delta_2)]$, $\delta_1=(\arcsin {\cal N}_{A-C})/2$, and $\delta_2=(\arcsin {\cal N}_{C-B})/2$ (clearly, this is only an example of the swapping-concentration schemes that can be performed: by properly choosing the basis for the measurement at location $C$, it is possible to have a trade-off between the final negativity of the new link $A-B$ and the success probability). If the swapping-concentration scheme has been successful, the use of this link provides a higher teleportation fidelity than the one that could be obtained by means of a double teleportation through the two original links $A-C$ and $C-B$ and can also outperform the one through the original link $A-B$. In this way, the average fidelity of teleportation has been increased. In the case where the swapping-concentration scheme has not been successful, both the state to teleport and the original link $A-B$ are unaffected, therefore one can still use the original path in order to achieve the maximum fidelity according to the formulas without the preparation stage.

The simple quantum scenario described here poses a very interesting problem. Once large quantum networks based on bi-partite entanglement are built up, their optimization during regular exploitation will require the development of new algorithms. Even though the optimal formulas we have derived can be beaten by allowing the protocol to become probabilistic, the violation of Bellman's optimality principle cannot be circumvented in the case where the probabilistic entanglement swapping for the nodes not belonging to the optimal route fails.

We thank M. S. Kim and M. Paternostro for discussions. C.D.F. was supported by Basque Government grant IT472-10, the Irish Research Council for Science, Engineering and Technology and UK EPSRC. D.B. acknowledges support from the SOLID European project and ESF.

\end{document}